\documentclass[
pra, 
aps,
twocolumn,nonatbib,
superscriptaddress,
longbibliography,
floatfix,
notitlepage]{revtex4-2}

\usepackage[export]{adjustbox}
\usepackage[utf8]{inputenc}
\usepackage[english]{babel}
\usepackage[T1]{fontenc}
\usepackage{lmodern}
\usepackage{epstopdf}
\usepackage{amsmath,amsfonts,amssymb,amsthm,bm,times,dcolumn,enumitem}
\usepackage[title]{appendix}

\usepackage{microtype}
\usepackage{braket}
\usepackage{gensymb} 
\usepackage{physics}

\usepackage[colorlinks={true}, citecolor={blue}, filecolor={blue}, linkcolor={blue}, urlcolor={blue}]{hyperref}
\usepackage{graphicx,color}
\usepackage[caption=false]{subfig}

\begin{document}

\preprint{APS/123-QED}

\title{Nanoscale addressing and manipulation of neutral atoms using electromagnetically induced transparency}

\author{U. Saglam}
\email[Electronic address:\ ]{usaglam@wisc.edu}

\author{T. G. Walker}
\email[Electronic address:\ ]{tgwalker@wisc.edu}
\author{M. Saffman}
\email[Electronic address:\ ]{msaffman@wisc.edu}
\author{D. D. Yavuz}
\email[Electronic address:\ ]{yavuz@wisc.edu}

\affiliation{Department of Physics, 1150 University Avenue, University of Wisconsin - Madison, Madison, Wisconsin 53706, USA}
\date{\today}

\begin{abstract}
We propose to integrate dark-state based localization techniques into a neutral atom quantum computing architecture and numerically investigate two specific schemes. The first scheme implements state-selective projective measurement by scattering photons from a specific qubit with very little cross talk on the other atoms in the ensemble. The second scheme performs a single-qubit phase gate on the target atom with an incoherent spontaneous emission probability as low as $0.01$. Our numerical simulations in rubidium (Rb) atoms show that for both of these schemes a spatial resolution at the level of tens of nanometers using near-infrared light can be achieved with experimentally realistic parameters. 
\end{abstract}

\maketitle

\section{\label{sec:intro}Introduction}

Over the last two decades, the interest in quantum computing has been continually growing due to the possibility of solving difficult computational problems efficiently~\citep{nielsen_chuang}. The principles of quantum computing have now been demonstrated using different physical qubits, each with various advantages and drawbacks, such as trapped ions~\citep{ionTrap}, superconducting qubits~\citep{superConducting1,superConducting2}, quantum dots~\citep{quantumDots1,quantumDots2}, nitrogen-vacancy centers~\citep{NVcenters}, and single photons~\cite{photonic}. In this paper, we will focus on addressing several challenges in neutral atom quantum computing. Neutral atoms have made great strides over the last decade towards a scalable quantum computing architecture~\citep{proposal_1, nielsen_chuang, saffman_Walker_review, proposal_4, proposal_5, proposal_6, proposal_7, proposal_8, proposal_9}. Single atoms can be trapped using microscopic dipole traps, and can be individually measured and addressed. Quantum information can be stored in the stable hyperfine states of the ground electronic level. Single qubit gates can be applied using microwave pulses~\cite{proposal_10,YWang2016,CSheng2018} or focused two-frequency Raman light~\cite{proposal_12,proposal_13}. Finally,  two-qubit gates are achieved by exciting the atoms to Rydberg states with a large principle quantum number (typically $n > 60$), and utilizing the dipole-dipole interaction~\cite{proposal_14}.

Scalability requires that gate errors are sufficiently low to be compatible with error correcting codes. Recent progress on both theory and experiment suggests that neutral atoms will be able to reach fidelity sufficient for error correction. For single qubit gates, experiments have already shown gate fidelities $>  0.9999$ \cite{CSheng2018}. For two-qubit Rydberg gates detailed theory has shown that fidelity $>0.999$ is possible, accounting for r atomic structure details and atomic recoil effects~\cite{proposal_36,proposal_37,Robicheaux2021,Pagano2022}. Although experiments are still far from the theoretical prediction, several groups have now demonstrated Rydberg state mediated entanglement with fidelity well above 90\% in long lived hyperfine ground 
states~\citep{proposal_42,neutral_new1,neutral_new2,ZFu2022}. In particular,  two-qubit gate fidelities exceeding 98\%\cite{ZFu2022} and a three-qubit Toffoli gate\cite{proposal_42} have been demonstrated, as well as 
multi-qubit entanglement and implementation of several quantum algorithms  using neutral atom arrays \cite{neutral_new1,neutral_new2}. Near term technical improvements in lasers, optical control, and cooling to reduce imperfections in Rydberg excitation, will likely lead to two-qubit gates and entanglement in the hyperfine basis, with a fidelity that is sufficient for error correction.

Despite this great progress, there are still outstanding challenges that need to be overcome in neutral atom quantum computing~\cite{saffman_Walker_review,proposal_44,Morgado2021}. Even with high fidelity gates,  implementation of cross-talk free qubit measurements (or qubit resetting~\cite{proposal_45}) which are required for error correction, is an outstanding challenge for neutral atom qubits. Hyperfine state selective projective measurements are typically performed by collecting fluorescent photons using a cycling transition. However, the photons scattered from a specific atom can be reabsorbed by neighboring atomic qubits causing errors as high as 4\% for realistic experimental conditions. This challenge can, in principle, be overcome by globally shelving all the other atoms in the array to a state not interacting with the fluorescent light~\cite{proposal_46,proposal_47}. However, this requires global operations on every qubit in the neighborhood of the measured atom, which is slow and adds to the error rate.

Another challenge in neutral atom arrays is the required high optical power of the trapping light. Due to the large overhead of quantum error correction, it is anticipated that from 100-1000 physical qubits may be needed for each protected logical qubit in a future universal computer. This implies that machines capable of beyond classical calculations based on several hundred logical qubits, may require $10^4$ to $10^5$ or more physical qubits. Present approaches based on two-dimensional arrays of optical traps with interatomic spacing of $d\sim 5 ~\mu\rm m$ require $>$100 Watts of optical trapping power to reach such large numbers.  The approach described in the following allows for spacing of $d\sim 0.5~\mu\rm m$, thereby reducing power requirements by a factor of 100, which will enable arrays with $>10^5$ qubits using only a few Watts of trapping light.

To address these challenges, we propose to integrate dark-state based localization techniques into a neutral atom quantum computing architecture. It is now well-understood that the dark state of electromagnetically induced transparency (EIT)  can be used to achieve a spatial resolution that can be much smaller than the wavelength of light. As we discuss below, the key idea is to use the spatial sensitivity of the dark state to the intensity of the coupling laser and tightly localize the coherent population transfer between the Raman levels. This approach was first theoretically proposed in Refs.~~\cite{Agarwal_2006, proposal_29_yavuzLab,proposal_30_lukin} and was experimentally demonstrated in \cite{diffBarOpticalShelwing,zubairyDiff,ZubairyScullyDarkState,proposal_31_yavuzLab,proposal_32_yavuzLab}. A related two dimensional scheme was proposed in Ref.~\cite{Wu_2014}. Very recently, EIT-based measurement in an ensemble without disturbing the quantum information has been considered as a promising method for error correction~\cite{weiss}. In this paper, we will focus on an array of qubits that are trapped in a standing-wave lattice, with a distance of $d \sim \lambda_{\rm lattice}/2=0.5$~$\mu$m between adjacent qubits. Due to their small spacing, it is difficult to address individual qubits using traditional optical means \cite{bloch_singlespin}. We will then discuss how variations of the dark-state-localization approach can be used to (i) perform state-selective projective measurement by scattering photons from a specific qubit with very little cross talk on the other atoms in the ensemble, (ii) implement a single-qubit phase gate with {\color{blue}low spontaneous emission rate} and with nanoscale position resolution.

The paper is organized as follows. In the model section, we discuss the dark state and EIT-based localization. In the results section, we will give details of the two schemes and show the results of the numerical simulations. In our simulations, we numerically solve the density matrix equations using realistic experimental parameters.

\section{\label{sec:model}Model}

\subsection{EIT and the Dark State}

Electromagnetically Induced Transparency (EIT) is a technique where the destructive quantum interference between different coupled energy levels of atoms are used to make a laser-dressed medium transparent to the probe light~\cite{proposal_66_imamoglu_rev, proposal_64_harris}. The absorption cancellation is because the atomic system is driven to the dark state which is a coherent superposition of the two lower Raman levels with no component in the excited radiating level. In addition to EIT, the dark state is also central to coherent population trapping (CPT) and stimulated-Raman adiabatic passage (STIRAP)~\cite{ImamogluRev_Lounis,ImamogluRev_CohenTannoudji}. The preparation of the dark state typically involves a pair of near-resonant fields both of which are coupled to an atomic lambda system. The Hamiltonian for the system can be written as $H=H_0+H_{\text{int}}$, where $H_0$ is the Hamiltonian of the bare state atom and $H_\text{int}$ describes the interaction between the atoms and the applied lasers.  The interaction Hamiltonian under the rotating wave approximation is:


\begin{figure}[h]
	\centering
	\subfloat[\label{fig:VeeConf}]{\includegraphics[scale=0.5]
		{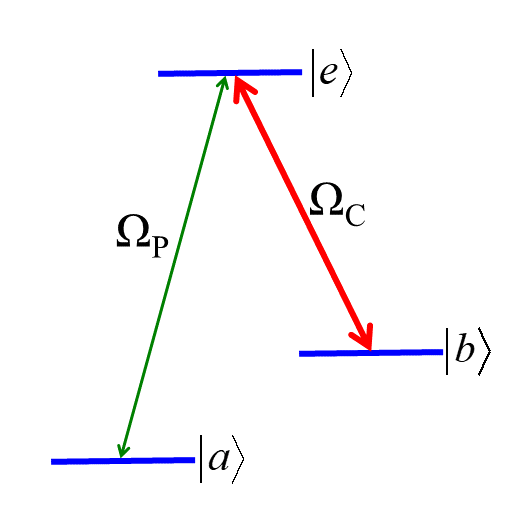}}
    \qquad
	\subfloat[\label{fig:generalScheme3}]{\includegraphics[scale=0.5]
		{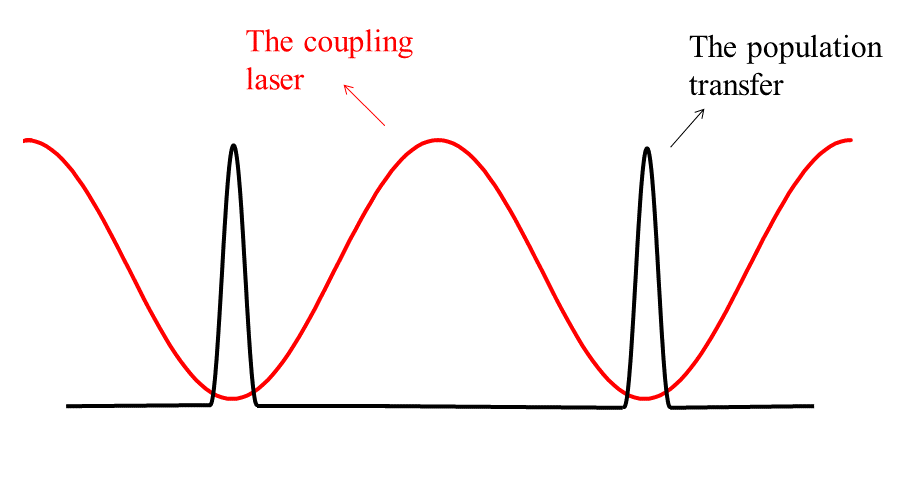}}
	\caption{\label{fig:genScheme}\textbf{(a)}~The three level $\Lambda$ configuration for EIT. Probe laser $\Omega_P$ couples level $\ket{a}$ with $\ket{e}$ and the coupling laser couples the level $\ket{b}$ to the level $\ket{e}$. \textbf{(b)}~Qualitative description of dark-state based localization of population transfer. With atomic system driven to the dark state, population transfer from the ground level $\ket{a}$ to $\ket{b}$ (black curve) can be localized to very small spatial scales. For comparison, the spatially varying coupling laser intensity (red curve) is also plotted.}		
\end{figure}


\begin{equation}
\label{eq:Hamiltonian}
H_{\text{int}}=
\begin{bmatrix}
0&0&\Omega_P\\
0&-2(\Delta_1-\Delta_2)&\Omega_C\\
\Omega_P&\Omega_C&-2\Delta_1\\
\end{bmatrix}.
\end{equation}

\noindent Noting Fig.~\ref{fig:VeeConf}, the quantities $\Omega_P$ and  $\Omega_C$ are the Rabi frequencies for the probe and the coupling lasers. The frequency detunings are defined as $\Delta_1=\omega_{e}-\omega_{a}- \omega_P$ and $\Delta_2=\omega_{e}-\omega_b-\omega_C$. The eigenvectors of the Hamiltonian of Eq.~(\ref{eq:Hamiltonian}), are

\begin{equation}
\label{eq:a+}
\ket{a^+}=\sin \theta \sin \phi \ket{a}+\cos \phi \ket{e}+\cos \theta \sin \phi \ket{b},
\end{equation}
\begin{equation}
\label{eq:a0}
\ket{a^0}=\cos \theta\ket{a}-\sin \theta \ket{b},
\end{equation}
\begin{equation}
\label{eq:a-}
\ket{a^-}=\sin \theta \cos \phi \ket{a}-\sin \phi \ket{e}+\cos \theta \cos \phi \ket{b}.
\end{equation}

\noindent where the quantities $\theta$ and $\phi$ are defined as:

\begin{equation}
\label{eq:eigTheta}
\tan \theta = \frac{\Omega_P}{\Omega_C} ,
\end{equation} 

\begin{equation}
\label{eq:eigPhi}
\tan 2\phi = \frac{\sqrt{\Omega_P^2+\Omega_C^2}}{\Delta_1}.
\end{equation}

The state $\ket{a^0}$  does not have any component in $\ket{e}$, and is the dark-state.
This state is smoothly connected to the ground state and can be prepared adiabatically using the counter-intuitive pulse sequence, i.e., the coupling laser beam turning on before the probe laser~\cite{ImamogluRev_Oreg, proposal_66_imamoglu_rev}. 
 In the dark state based localization approach, the key idea is to use the spatial sensitivity of the dark state to the intensity of the coupling laser beam. From Eq.~(3), it can be shown that the population of state $|b \rangle$ is $|\langle b | a^0 \rangle |^2 = |\Omega_P|^2/(|\Omega_P|^2+|\Omega_C|^2$), and has a highly nonlinear dependence on coupling laser intensity. Specifically, in a region where the coupling laser goes through an intensity minimum,  the population of state $|b \rangle$ can be localized very tightly. As discussed in Refs.\cite{proposal_30_lukin,proposal_32_yavuzLab}, an easy approach to implement this scheme would be to use a coupling laser beam with a standing-wave spatial profile. This is schematically shown in  Fig.~\ref{fig:generalScheme3}.  It can be shown that, for a spatially uniform probe laser, the full-width-half-maximum of the transfer will be, FWHM$_{transfer} \approx \lambda (\Omega_P/\Omega_{C,max})$, where $\Omega_{C,max}$ is the Rabi frequency of the coupling laser at the peak of the standing wave.

Using the dark state for the localization of population transfer provides key advantages compared to other approaches ~\cite{proposal_68,proposal_69,proposal_70,proposal_71,proposal_72,proposal_73,proposal_74,proposal_75,proposal_76,proposal_77,proposal_78,proposal_79,proposal_80,proposal_81,proposal_82,proposal_83,proposal_29_yavuzLab}. The atoms are coherently transferred, keeping their phase relationship with other qubits intact. If the evolution is sufficiently adiabatic, the dark state can be prepared with little population in the excited radiative state, which reduces heating and decoherence from spontaneous emission. Because the excitation is coherent, dark-state-based localization can be achieved using short and intense laser pulses. Finally, due to the robust nature of adiabatic preparation, the localization is insensitive to fluctuations in experimental parameters, such as frequency and intensity jitters of the probe and coupling lasers. The protocols that we discuss below for state-selective readout and single-qubit phase-gate fully utilize these advantages


\begin{figure}[h]
	\centering
	\includegraphics[width=\linewidth]{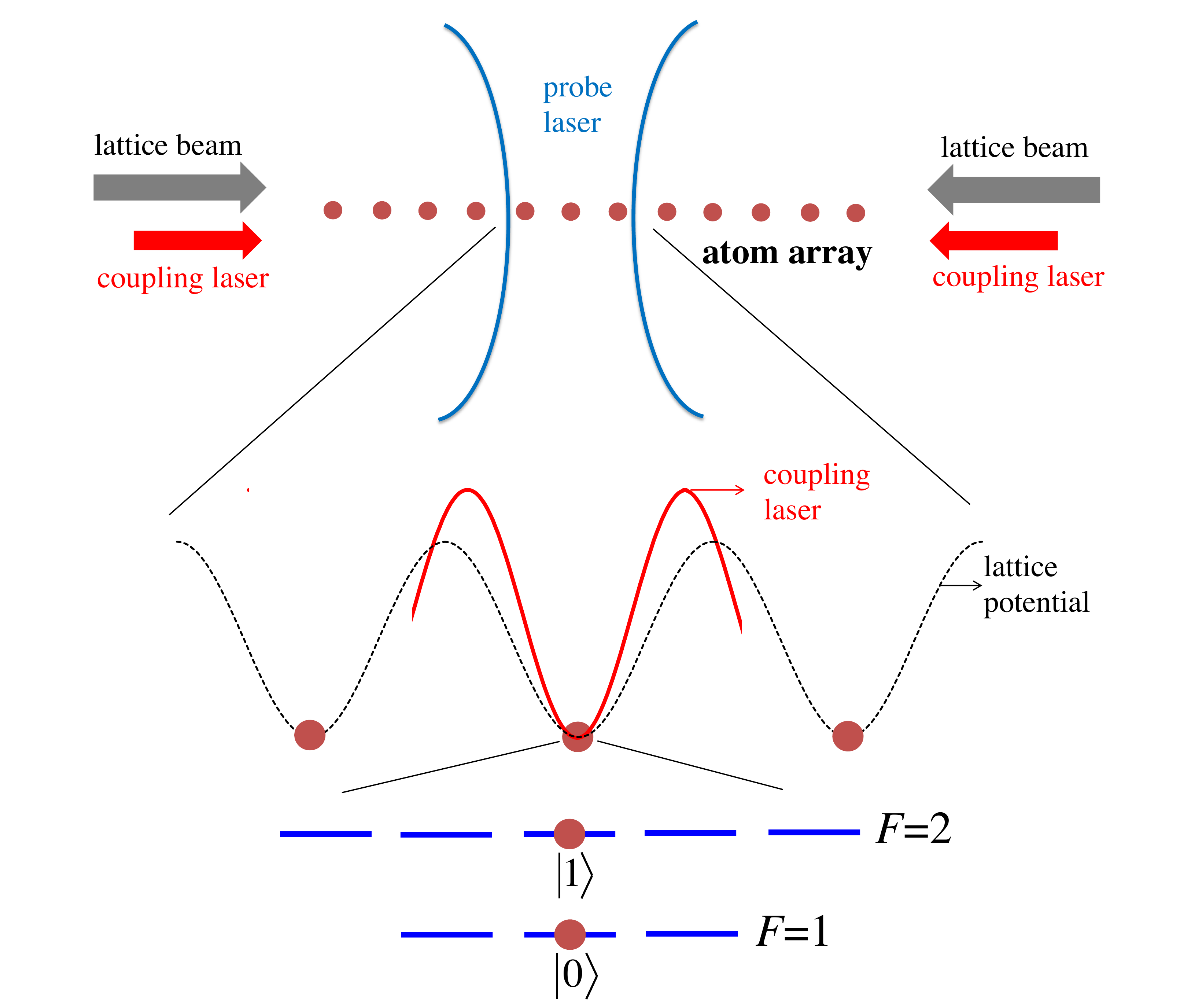}
	\caption{\label{fig:generalScheme_Lattice}  Nanoscale measurement and addressing scheme. The atoms are trapped in a one-dimensional optical lattice which is obtained using a counter-propagating beam pair(the lattice potential is shown in dashed black curve). The intensity minimum of the coupling laser of EIT (solid red curve) is aligned to the qubit of interest. The coupling laser standing-wave is also obtained using a counter-propagating beam pair. The probe laser beam is focused sufficiently tightly so that it overlaps with only the three qubits in the array (i.e., to a spot size of $\sim \lambda_{lattice}=1.17$~$\mu$m). In our simulations, we focus on nanoscale addressing and manipulation of the central qubit with negligible cross-talk to the neighboring two qubits.  The logical qubit states are the clock states of the ground hyperfine manifold:  $|0 \rangle \equiv |F=1, m_F=0 \rangle$ and $|1 \rangle \equiv |F=2, m_F=0 \rangle$.}
\end{figure}

\section{Results}
\label{section:results}

Figure~2 shows the specific system that we will be focusing on throughout the rest of the paper. We consider a neutral atom based quantum computing architecture using $^{87}$Rb atoms and take clock states as the logical qubit states: $|0 \rangle \equiv |F=1, m_F=0 \rangle$ and $|1 \rangle \equiv |F=2, m_F=0 \rangle$. For simplicity and clarity, we focus on a one-dimensional geometry, although the protocols that we discuss extend to two-dimensions in a straightforward way. The atoms are trapped in a lattice using a far-off-resonant dipole trap.  Nanoscale state-selective measurement and single-qubit gates are achieved using a spatially varying coupling laser beam at a wavelength near the D2 line of $\lambda_{\rm D2}=780$~nm. The coupling laser standing-wave is obtained by using a counter-propagating beam-pair. The intensity minimum of the coupling-laser beam is interferometrically aligned to the qubit that is to be addressed. We choose the wavelength of the dipole trap laser such that the coupling laser is at a maximum at the nearby qubits: $\lambda_{\rm lattice} = \frac{3}{2} \lambda_{\rm D2} = 1.17$~$\mu$m. This results in a qubit spacing of $\lambda_{\rm lattice}/2=  0.59$~$\mu$m. To simplify the discussion, we will take the probe laser beam to be focused sufficiently tightly so that it overlaps with only the three qubits in the array (i.e., to a spot size of $\sim \lambda_{lattice}=1.17$~$\mu$m). In what follows, we will focus on nanoscale addressing and manipulation of the central qubit with negligible cross-talk to the neighboring two qubits.  

\subsection{State-selective single-qubit readout }

As mentioned above, state selective single-qubit readout remains one of the outstanding challenges in neutral atom quantum computing. This is difficult to achieve even in arrays where the atoms are spaced by $\sim$10 microns and individual addressing is achieved by tightly focused laser beams. State-selective projective measurement is traditionally performed by collecting fluorescent photons using a cycling transition. However the photons scattered from a specific atom in the array can be reabsorbed, thereby causing error rates as high as 4\% at neighboring qubits $\sim 10$~microns away. This challenge can be overcome by globally shelving remaining qubits in the array to a state not interacting with the fluorescent light~\citep{saffman_Walker_review,proposal_44}. However, this then requires global operations on every qubit in the ensemble, which is slow and adds to the error rate. In this section, we discuss a dark-state localization based measurement scheme that largely overcomes these challenges even when the spacing between adjacent qubits is 0.59~$\mu$m. 

\begin{figure}[h]
	\label{fig3}
	\centering
	\includegraphics[width=\linewidth]{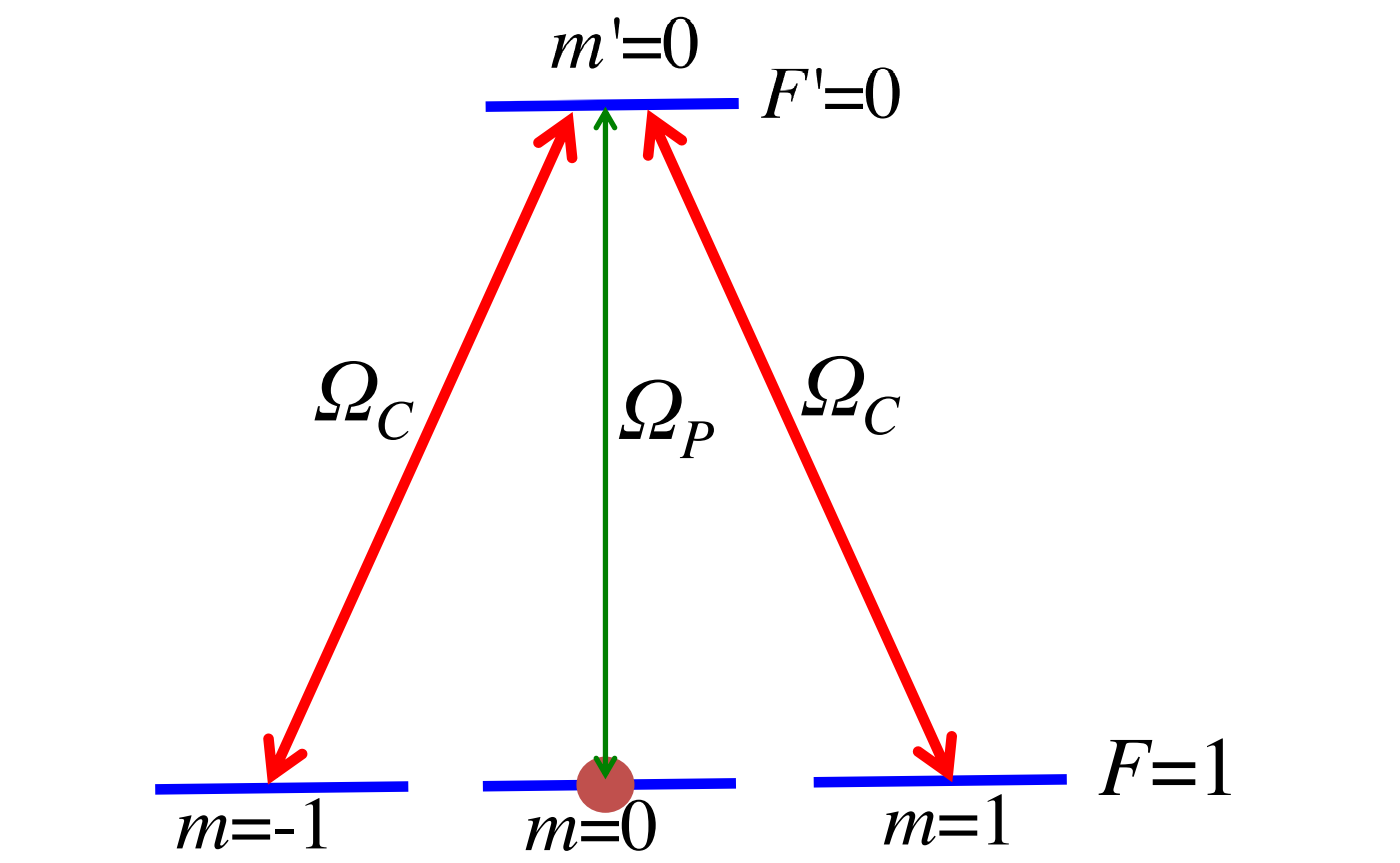}
	\caption{\label{fig:fourLevel} Level scheme in $^{87}$Rb for nanoscale-level single-qubit readout, which consists of two parallel EIT channels. The coupling laser with Rabi frequency $\Omega_C$ couples the states $\ket{F=1,m_F=\pm1}$ and $\ket{F'=0,m_{F'}=0}$ and the probe laser with Rabi frequency $\Omega_P$ couples the states $\ket{F=1,m_F=0}$ and $\ket{F'=0,m_{F'}=0}$.}
\end{figure}
\noindent

\hfill

Figure~\ref{fig:fourLevel} shows the relevant energy level diagram for the measurement scheme that we envision. The goal is to perform a projective measurement of the logical $|0 \rangle$ state of only the addressed qubit, while scattering as few photons from the other qubits as possible.  A probe laser beam polarized along the quantization axis ($\pi$ polarization) couples $|F=1, m_F=0 \rangle$ to $|F'=0, m_{F'}=0 \rangle$ state of the D2 line. The two beams forming the coupling laser standing wave are linearly polarized orthogonal to the quantization axis, thereby containing equal amounts of $\sigma^+$ and $\sigma^-$ light. As shown in Figure~\ref{fig:fourLevel}, the result is two $\Lambda$ schemes, forming two parallel EIT channels. 

\begin{figure}[h]
	\centering
	\includegraphics[width=\linewidth]{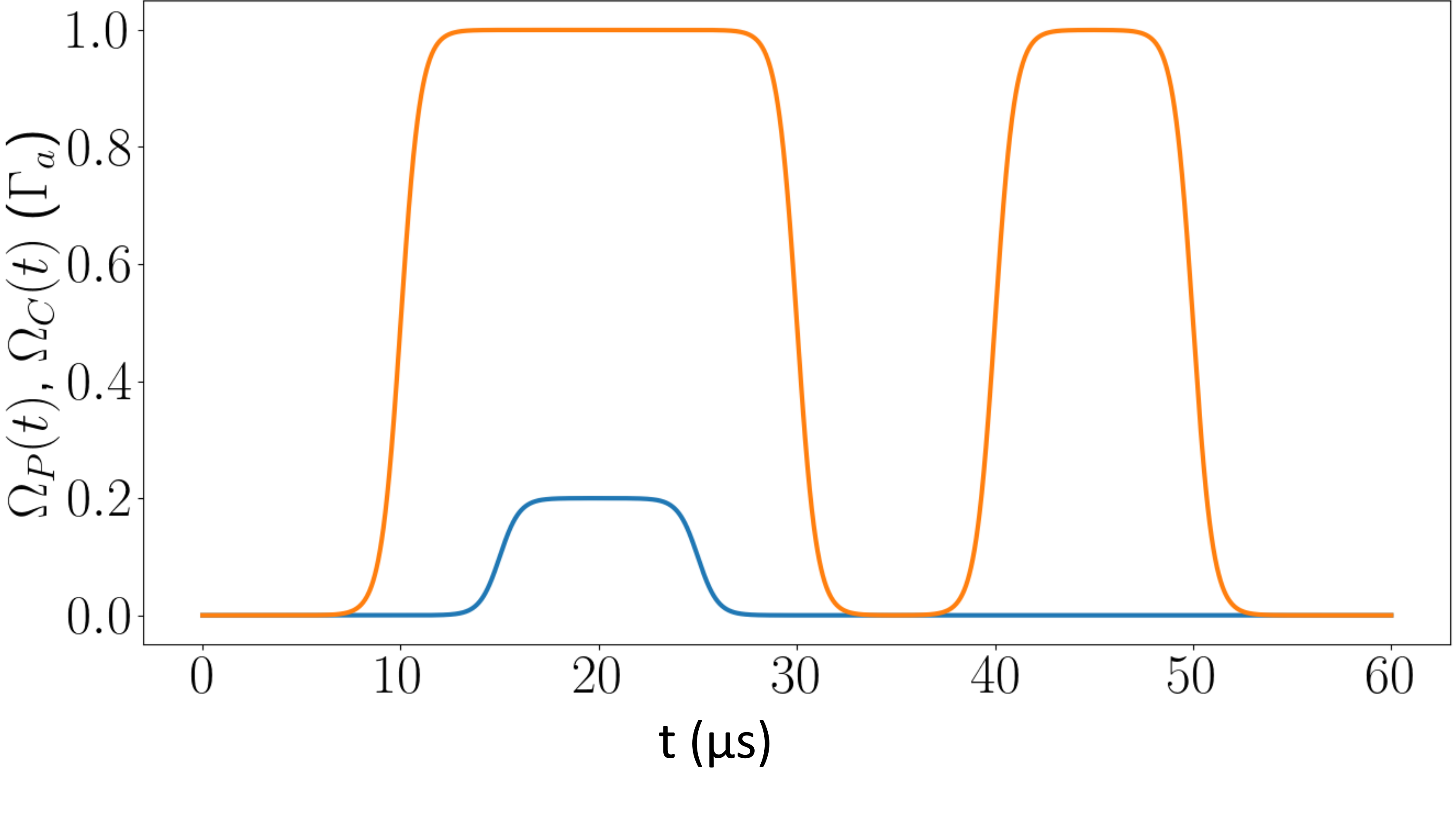}
	\caption{\label{fig:readout} The Rabi frequencies for the coupling laser $\Omega_C$, (solid, orange curve) and probe laser $\Omega_P$, (solid, blue curve) as a function of time. The lasers are turned on adiabatically using a counter-intuitive pulse sequence: i.e., the coupling laser is turned on before the probe laser. Second coupling-only pulse is for optically pumping the atom back to the logical $\ket{0}$ state. }
\end{figure}

The measurement protocol is as follows. During the EIT pulse, the intensities of the beams forming the coupling laser standing-wave are balanced so that the intensity vanishes at the minimum. At the node of the coupling laser (which is the position of the qubit whose state is to be measured), the atom scatters probe photons and is pumped into one of $|F=1, m_F=1 \rangle$ or $|F=1, m_F=-1 \rangle$ states (since EIT is not established). The other atoms in the array experience EIT and evolve into the dark state. As shown in  Fig.~\ref{fig:readout}, the coupling laser is turned on before and turned-off after the probe beam. As a result, the other qubits in the array adiabatically evolve from the $|0 \rangle$ state into a coherent superposition and then evolve back to $|0 \rangle$. This is achieved while maintaining negligible population in the excited $|F'=0, m_{F'}=0 \rangle$ state  and therefore with low spontaneous emission. After the EIT (probe and coupling) pulse sequence, the atom at the node is left in one of the $|F=1, m_F=1 \rangle$ or $|F=1, m_F=-1 \rangle$ states. To pump this atom back, we turn on the coupling laser beam, but now with a slight intensity imbalance in the beam-pair, so that there is some light at the intensity minimum. Note that this second coupling-only pulse does not interact with the other atoms in the array (since all the other atoms are in state $|0 \rangle$). The end result after this pulse sequence is that the atom at the node scatters $\sim$2 photons, while other atoms in the array scatter little if efficient EIT is achieved. To ensure arrival of a photon at the detector, the above pulse sequence can be applied multiple times to scatter a sufficient number of photons from the addressed qubit.

One of the key advantages of this approach is that due to the presence of EIT at the other atoms, there is negligible probability of scattered photons to be reabsorbed within the array. If the coupling laser intensity is much larger than the probe laser intensity at the positions of the other atoms, the majority of the dark state remains in state $|0 \rangle$. As a result, the probability of reabsorption is significant only for photons scattered on the probe transition ($\pi$ polarized). But these photons are scattered only when there is large coupling intensity at the other atoms, which means the atoms are transparent to these photons due to EIT.

In this approach, the crosstalk and the error on the adjacent qubits is limited by non-adiabatic corrections to the dark state. This can be kept quite low, by ensuring that $(\delta \omega/\Omega_C)^2$ (the quantity $\delta \omega$ is the bandwidth of the probe pulse) is sufficiently small at the neighboring qubits. However, $\delta \omega$ cannot be set arbitrarily low, since the bandwidth of the probe pulse essentially determines the measurement time.

%

\hfill

\hfill

\begin{figure*}[t]
	\label{fig5}
	
	\includegraphics[width=\linewidth]{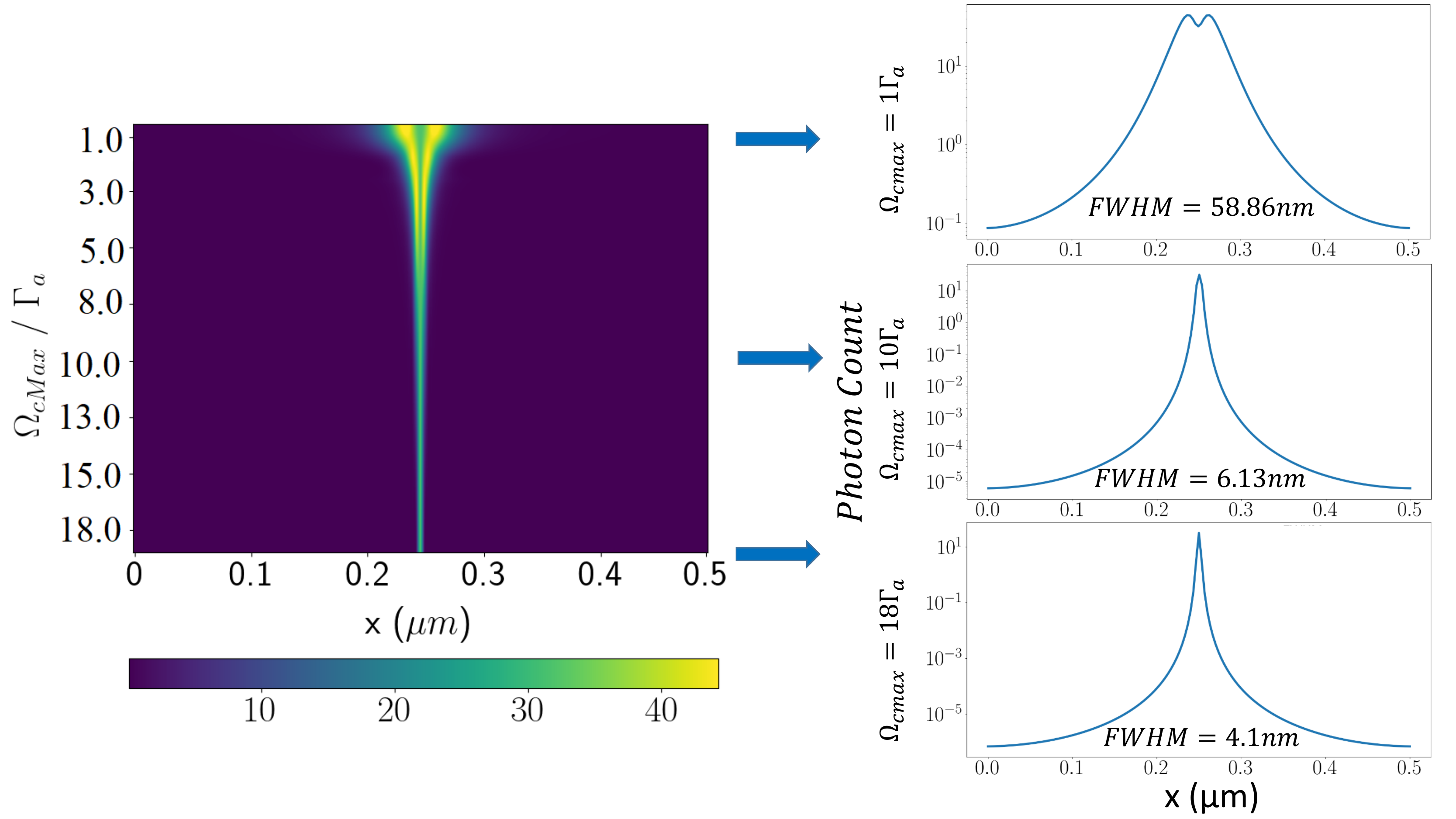}
	\caption{\label{fig:LocalizationCrossSection}Dependence of the localization with respect to the maximum coupling laser frequency $\Omega_{C,\text{max}}$ and the relevant cross sections of the 2D plot with $\Omega_{C,\text{max}}=1\Gamma_a$, $\Omega_{C,\text{max}}=10\Gamma_a$, $\Omega_{C,\text{max}}=18\Gamma_a$. With FWHMs of 58.5~nm's, 6.1~nm's 4.1~nm's, respectively. }
\end{figure*}

\hfill

Figure~\ref{fig:LocalizationCrossSection} shows the results of a simulation where we numerically solve the $4 \times 4$ density matrix for the scheme of Fig.~\ref{fig:fourLevel} using quite reasonable parameters. The equations for the time evolution of the density matrix, as well as their derivation is outlined in Appendix~A. In these simulations, we take probe pulses with a duration of 6~$\mu$s, a probe Rabi frequency of $\Omega_P=0.2 \Gamma_a$ (the quantity $\Gamma_a=2 \pi \times 6.06$~MHz is the D2 line decay rate). In the false-color two-dimensional plot, the coupling laser Rabi frequency at the peak of the standing wave is varied from $\Omega_{C,max}=\Gamma_a$ to $\Omega_{C,max}=18 \Gamma_a$.  
We take a combined photon collection efficiency of 3\% (a numerical aperture of NA=0.5 of the initial lens and 40\% detection efficiency from the first lens to the photon counter). Figure~\ref{fig:LocalizationCrossSection} shows the total number of scattered photons (in log scale) as a function of position. The atom at the node scatters 33 photons (which produces a mean detected photon number of 1), and this scattering is spatially localized very strongly. The three insets show the number of scattered photons as a function of position for $\Omega_{C,max}=\Gamma_a$, $\Omega_{C,max}=10 \Gamma_a$, and $\Omega_{C,max}=18 \Gamma_a$, respectively. For these three cases, The population transfer is localized to a spatial region with a FWHM of 58.5~nm, 6.1~nm, and 4.1~nm, respectively. Even tighter localization of photon scattering can be achieved with the use of higher values for the coupling laser Rabi frequency at the peak of the standing wave.

We note that, scattering 33 photons requires $\sim 33/2 \sim 16$ pulse sequences. This sets the measurement time for the numerical simulations of Fig.~~\ref{fig:LocalizationCrossSection} to be $\sim 16 \times 6 \sim 100$~$\mu$s. Faster measurement times can be achieved with the use of shorter pulses, which would increase the bandwidth, $\delta \omega$, resulting in higher nonadiabatic corrections to the dark state (and therefore higher crosstalk) for fixed probe and coupling laser Rabi frequencies. This can be overcome by increasing
the probe and coupling laser Rabi frequencies while keeping their ratio constant (in order to achieve a similar amount of spatial localization).

\subsection{The effect of initial spread of the atomic position}

The numerical simulations of Fig.~5 assume the ideal case of no spread of the initial atomic position (i.e., the atom is assumed to be a point particle at a fixed position). In a realistic experiment, there will be an initial spread of the atomic position due to the finite depth of the optical trap and the atomic temperature. This initial spread of the atomic position will broaden the results that are presented in Fig.~5. We calculate this broadening to be on the order of tens of nanometers, with trap depth of the lattice potential in the 1-10~mK range and atoms cooled to the ground state of the trap (with the wavelength of the lattice potential fixed at $\lambda_{lattice}=1.17$~$\mu$m, and therefore the qubit spacing fixed at $\lambda_{lattice}/2$=0.59~$\mu$m).

Figure~6 shows our numerical calculation of this effect for the specific case of a trap depth of 5~mK. Here, we assume the atom to be in the ground state of the trapping potential, calculate the initial probability spread for the atomic position, and convolve the result that we obtain in Fig.~5 with this initial probability distribution. The result of Fig.~6 is obtained for the case of $\Omega_{C,max}=18\Gamma_a$, i.e., for the conditions of the third inset in the numerical simulation of Fig.~5. The FWHM spread for the scattered photons is increased to $\sim$$23.8$ nm (from  4.1~nm). This is still well below the diffraction limit of the addressing probe and coupling lasers, which has a wavelength of $\lambda_{D2}=780$~nm.

\begin{figure}[htb]
    \centering
    \includegraphics[width=\linewidth]{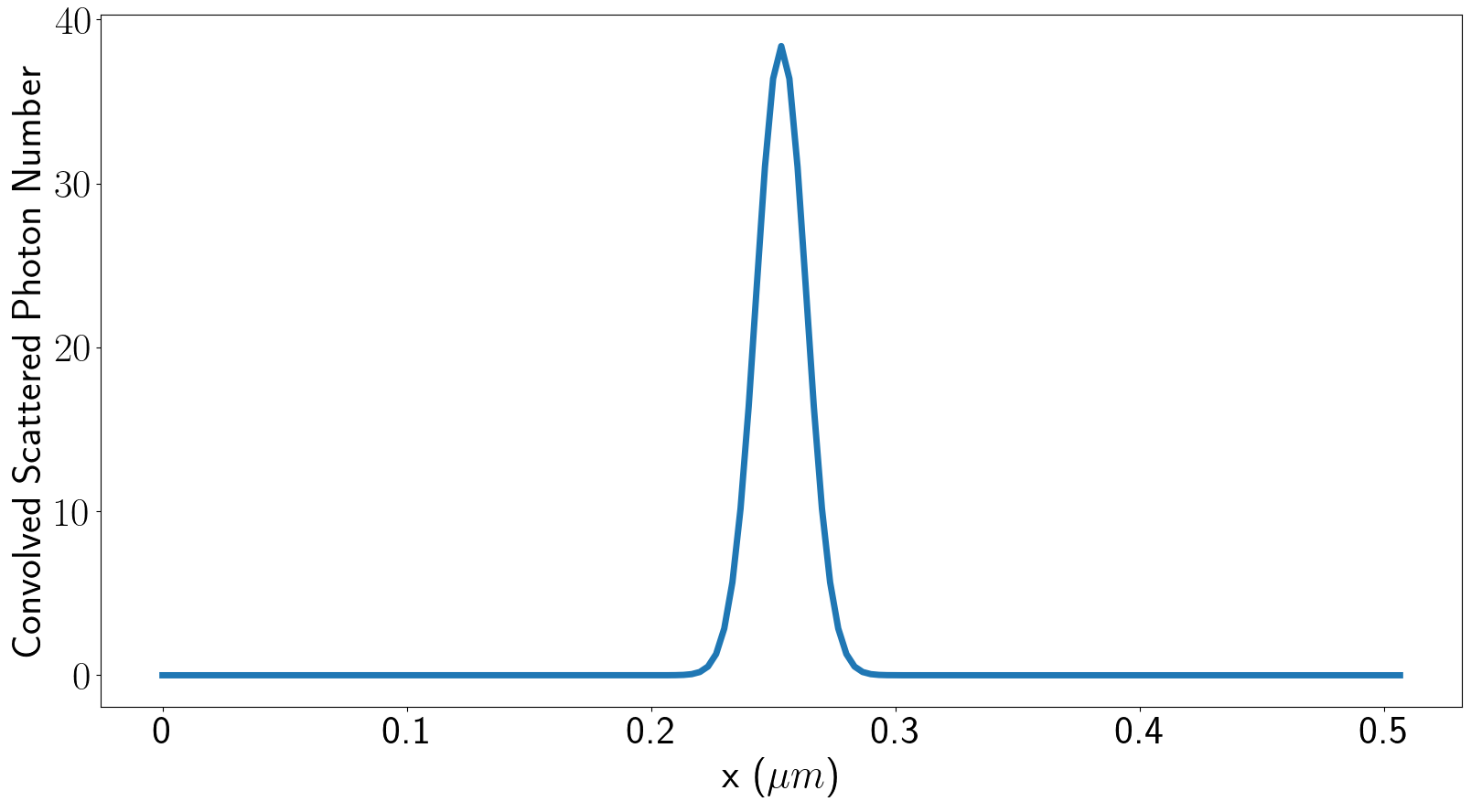}
    \caption{\label{fig:conv}The photon count plot for $\Omega_{C,max}=18\Gamma_a$ when we assume the atom to be in the ground state of a trapping potential with the trap depth of 5~mK. We obtain this result by convolving the numerical calculation for the third inset of Fig.~5, with the probability distribution of the atomic position due to the finite spread of the wavefunction. The FWHM of the localization for the scattered number of photons is increased to $23.8$~nm. }
    \label{fig:my_label}
\end{figure}

\subsection{The effect of photon scattering on the nearby qubits}

In the simulations of Fig.~5, we numerically calculated the density matrix equations due to the probe and coupling lasers only. These results show that within the assumptions of these simulations, the addressed qubit can scatter many photons while keeping the scattering from the nearby qubits at very low values. We have also qualitatively argued that the photons scattered  from the the targeted atom will not be reabsorbed by the nearby atoms due to the presence of EIT. In this section we will make this argument quantitative by including the effect of the radiated field from the addressed qubit on the nearby atoms explicitly. For this purpose, we first calculate the electric field at the position of the neighboring qubit, due to the scattered photons from the targeted atom. Consider a radiating dipole (the target qubit), with a dipole moment, $\vec{p}$. Working in a spherical coordinate system and positioning the addressed atom at $r=0$, the vectorial electric field due to this radiating atom is:
\begin{eqnarray}
\vec{E}_{dipole} & = & \frac{k^3}{4 \pi \epsilon_0} \{ \frac{1}{k r } ( \hat{r} \times \vec{p} ) \times \hat{r} \quad \nonumber \\
& + &  \left( \frac{1}{k^3 r^3} - i \frac{1}{k^2 r^2} \right) [3 \hat{r} (\hat{r} \cdot \vec{p}) - \vec{p} ]  \} \quad . 
\end{eqnarray}

\noindent Here, the quantity $\hat{r}$ is the unit vector that connects the observation point to the radiating atom and $k = 2 \pi /\lambda$ is the optical $k$-vector. As the targeted atom is radiating, depending on which specific transition the excited atom decays into, the orientation of the dipole vector $\vec{p}$ can be different. A photon that is emitted into the $| F'=0, m_{F'}=0 \rangle  \rightarrow | F=1, m_{F}=0 \rangle$ transition will be linearly polarized while a photon emitted into the $| F'=0, m_{F'}=0 \rangle  \rightarrow | F=1, m_{F}=\pm 1 \rangle$ transitions would be circularly polarized. Scattered photons at these different polarizations would then introduce perturbations to the probe and coupling laser frequencies at the position of the neighboring qubit. 
We calculate these time varying perturbations, which we refer to as $\Omega_{P,dipole}(t)$ and $\Omega_{C,dipole}(t)$, by multiplying the radiated electric field at the position of the neighboring qubit with respective matrix elements of the transitions. 

We have simulated this effect by adding the time dependent dipole emission to the density matrix equations at the nearby qubit which is $r=0.59$~$\mu$m away. At this distance the peak values for the perturbation to the Rabi frequencies are $\Omega_{P,dipole} = 2 \pi \times 159$~kHz and $\Omega_{C,dipole} =  2 \pi \times 159$~kHz, respectively. With these added dipole perturbations to the Rabi frequencies for the probe and coupling laser beams, we then calculate the change in the photon scattering rate at this other qubit, when compared with the simulations of Fig.~5. We find that the results that we have reported in Fig.~5 only change at the level of $10^{-6}$ or less at the position of the adjacent neighboring qubit. This is remarkably low and is due to the robustness of EIT to the exact values of the probe and coupling laser intensities.

\begin{figure}
	\label{fig1}
	\centering
	\includegraphics[width=\linewidth]{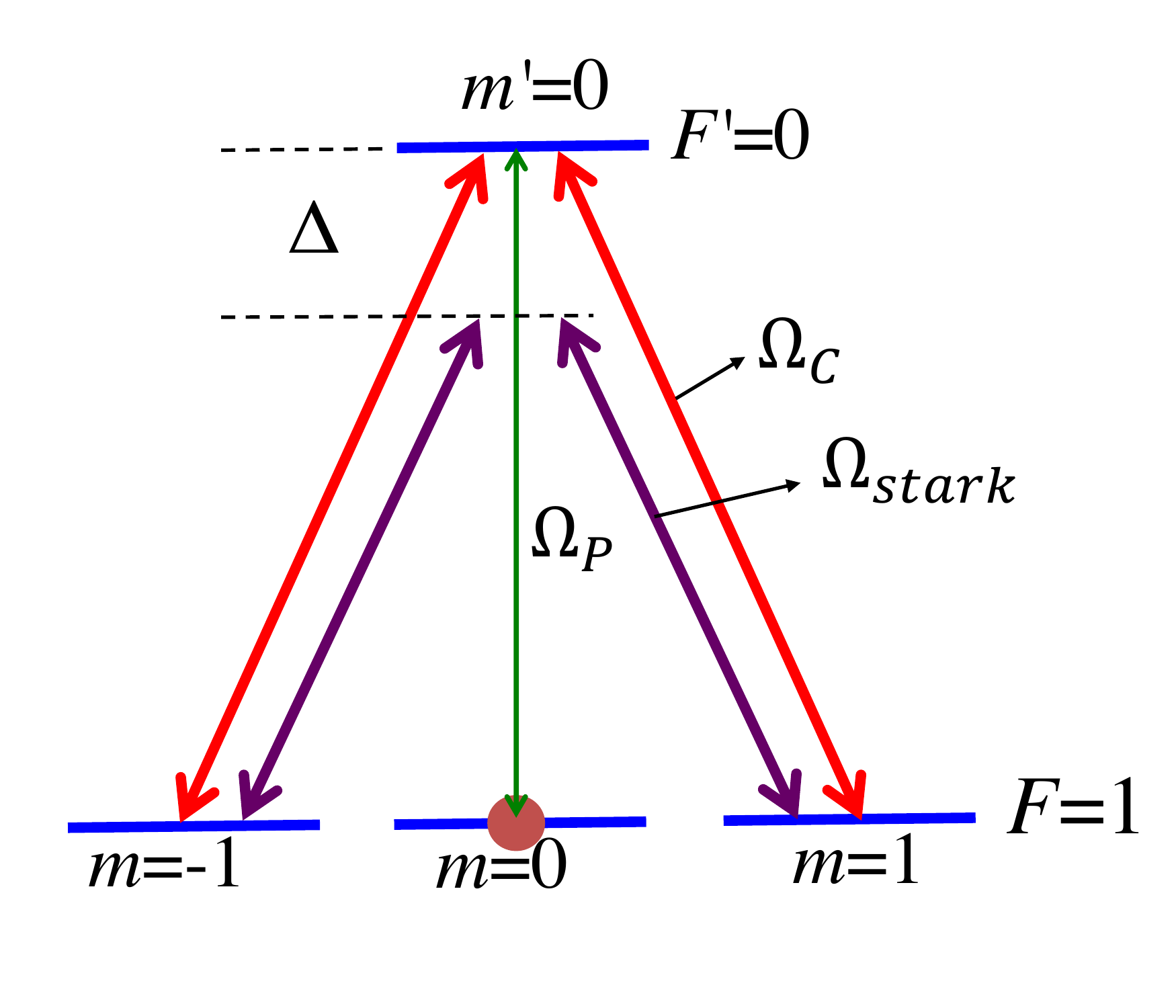}
	\caption{\label{fig:generalSchemePhase}  Four level structure $F'=0, m_F=0$ and $F=1, m_F=0,\pm 1$with the relevant coupling and probe lasers, $\Omega_C$ and $\Omega_P$. In addition, there is a detuning of $\Delta$ and new Stark-Shift Laser with Rabi Frequency $\Omega_{stark}$. }
\end{figure}

\begin{figure}
	\label{fig4}
	\centering
	\includegraphics[width=\linewidth]{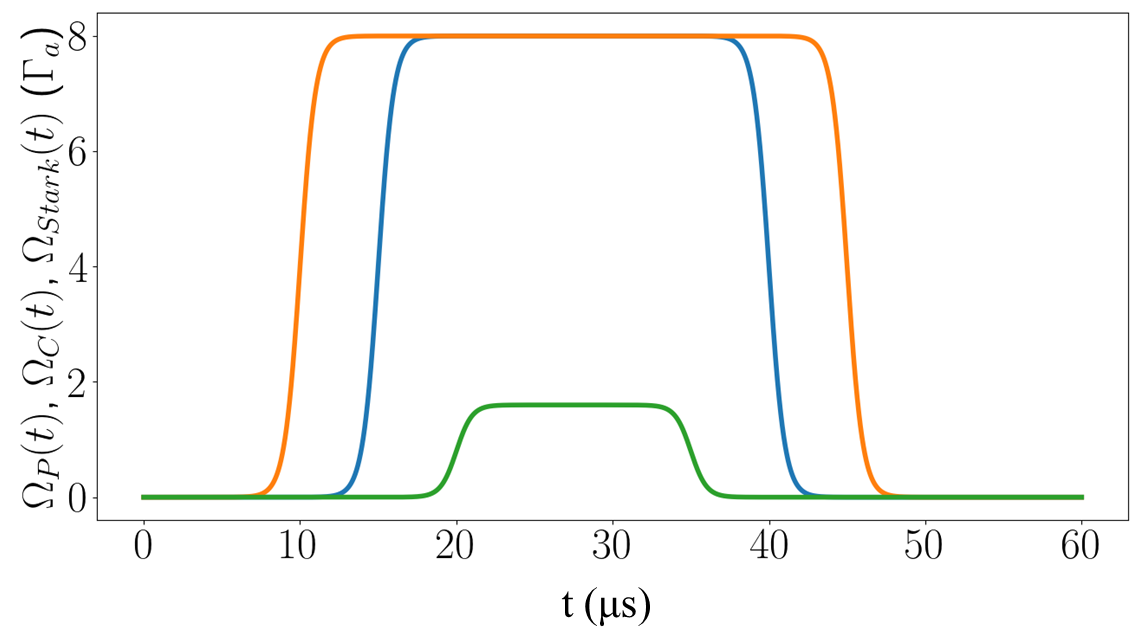}
	\caption{\label{fig:PhaseGateFreq} Plot of the Rabi frequencies for the phase gate scheme with the minimal spatial dependence, $\Omega_{C,max}=8\Gamma_a$. All of the frequencies are turned on adiabatically, where the turn order is $\Omega_C$ (Solid, orange curve), $\Omega_P$ (Solid, blue curve) and $\Omega_{stark}$ (Solid, green curve)}
\end{figure}

\begin{figure*}
	\includegraphics[width=\linewidth]{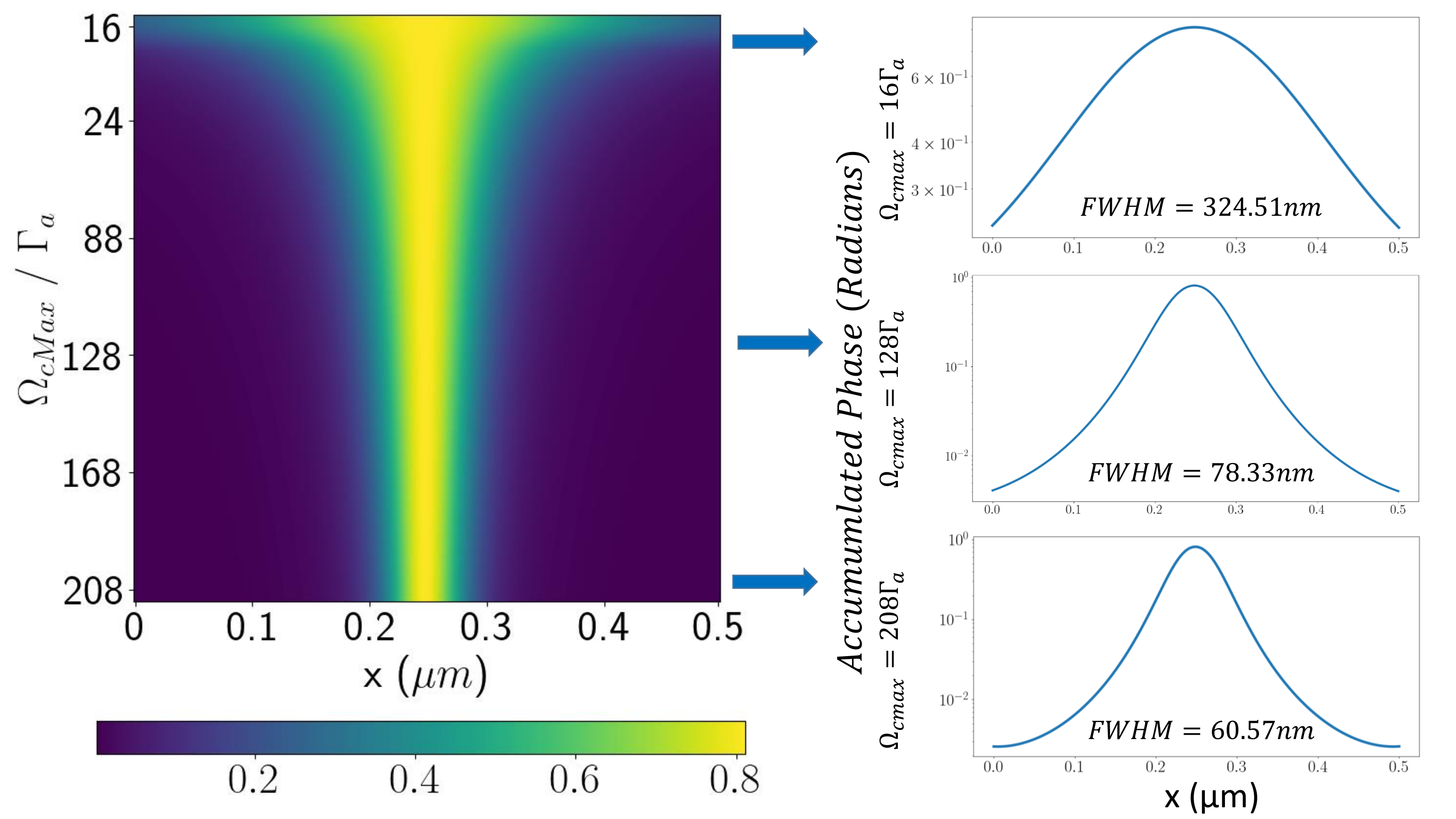}
	\caption{\label{fig:PhaseCrossSection}Dependence of the localization of the phase with respect to the maximum coupling laser frequency $\Omega_{C,\text{max}}$ and the relevant cross sections of the 2D plot with $\Omega_{C,\text{max}}=16\Gamma_a$, $\Omega_{C,\text{max}}=128\Gamma_a$, $\Omega_{C,\text{max}}=208\Gamma_a$. With FWHMs of 324.51~nm's, 78.33~nm's, 60.57~nm's, respectively. }
\end{figure*}

\subsection{Single-qubit Phase Gate}

In this section we discuss how to implement a single-qubit phase gate with the truth-table $|1\rangle \rightarrow |1\rangle$, $|0\rangle \rightarrow \exp{(i \varphi)} |0\rangle$. The phase gate, although not universal by itself, does provide for arbitrary single qubit rotations on targeted qubits when combined with global $\pi/2$ and $-\pi/2$ rotations about an axis on the equator of the Bloch sphere which can be readily implemented with microwaves\cite{YWang2016,neutral_new1}.

Our approach is similar to what was discussed in Ref.~\cite{proposal_30_lukin}. The relevant energy level diagram and the pulse sequences for the involved laser beams are shown in Figs.~7 and 8. To implement a single-qubit phase-gate, we are interested in coherent manipulation of the atom near the node of the coupling laser spatial profile using EIT. For this purpose, similar to the localization scheme that we discussed above, we need to ensure adiabatic evolution at all points along the coupling-laser standing wave. We therefore introduce an imbalance in the intensities of the beam-pair forming the coupling laser, which then results in a non-vanishing intensity minimum. If the peak coupling laser intensity is much higher than the uniform probe intensity, then there is localization of the transfer to states $|F=1, m_F=1 \rangle$ and $|F=1, m_F=-1 \rangle$. Only near the intensity minimum of the coupling laser profile, there is large population transfer from the logical $|0\rangle$ state to these states during EIT. We then adiabatically turn on a far-detuned laser beam, which Stark-shifts the  $|F=1, m_F=1 \rangle$ and $|F=1, m_F=-1 \rangle$ states by an amount $\Omega_{stark}^2/(2 \Delta) $. The Stark-shift causes a phase accumulation of $\Omega_{stark}^2/(2 \Delta)  T $ (the quantity $T$ is the duration of the Stark-shift laser pulse). But only near the intensity minimum where there is substantial transfer to the $|F=1, m_F=1 \rangle$ and $|F=1, m_F=-1 \rangle$ states, the wavefunction acquires this phase.

The Rabi frequencies for the laser pulses as a function of time are plotted on Fig.~\ref{fig:PhaseGateFreq}. To ensure adiabatic preparation, the coupling laser beam is turned-on before the probe laser beam. The Stark-shift laser is present only after EIT has been established and the system has evolved into the dark state (i.e., after the probe and coupling laser beams are turned-on).

Fig.~\ref{fig:PhaseCrossSection} shows the result of density-matrix numerical simulations for the single-qubit phase-gate scheme. Here, the maximum of the coupling laser Rabi frequency is varied to until $\Omega_{C,\text{max}}=208\Gamma_a$, while the value of the coupling laser Rabi frequency at the intensity minimum is kept fixed at $\Omega_{C,\text{min}}=8\Gamma_a$.  We set the uniform probe laser Rabi frequency to be $\Omega_P=8\Gamma_a$. The parameters of the Stark shift-laser beam are adjusted to obtain a phase-shift value of $\pi/4$ radians. Specifically, we choose the Rabi frequency of the Stark-shift laser to be $\Omega_{stark}=1.6\Gamma_a$, set it's detuning at $\Delta = 200 \Gamma_a$, and take it's duration to be $T=15\mu s$. The durations of the probe and coupling laser pulses as shown in Fig.~7 are $25 \mu s$ and $35 \mu s$, respectively. 

The 2D plot in Fig.~\ref{fig:PhaseCrossSection} is a false-color plot of the applied phase as the Rabi frequency of the coupling laser at the intensity maximum, $\Omega_{C,\text{max}}$ is varied. The three insets show the number of scattered photons as a function of position for $\Omega_{C,max}=16\Gamma_a$, $\Omega_{C,max}=128 \Gamma_a$, and $\Omega_{C,max}=208 \Gamma_a$, respectively. For these three cases, The population transfer is localized to a spatial region with a FWHM of 324.5~nm, 78.3~nm, and 60.6~nm, respectively.

In this scheme, the fidelity of the single-qubit phase-gate is limited by spontaneous emission (and therefore decoherence) of the excited radiating level, $|F'=0, m_{F'}=0 \rangle$. This level is populated due to (1) non-adiabatic corrections to the dark state, and (2) the far-detuned excitation because of the Stark-shift laser. As discussed above, the non-adiabatic corrections to the dark state can be kept low by keeping the ratio of the bandwidth of the probe and coupling laser pulses to their Rabi frequencies to be small. The excitation of the radiating level due to the Stark-shift laser can be kept small, by keeping a large value for the detuning of the Stark-shift laser beam, $\Delta$.

\begin{figure}
	\label{fig10}
	\centering
	\includegraphics[width=\linewidth]{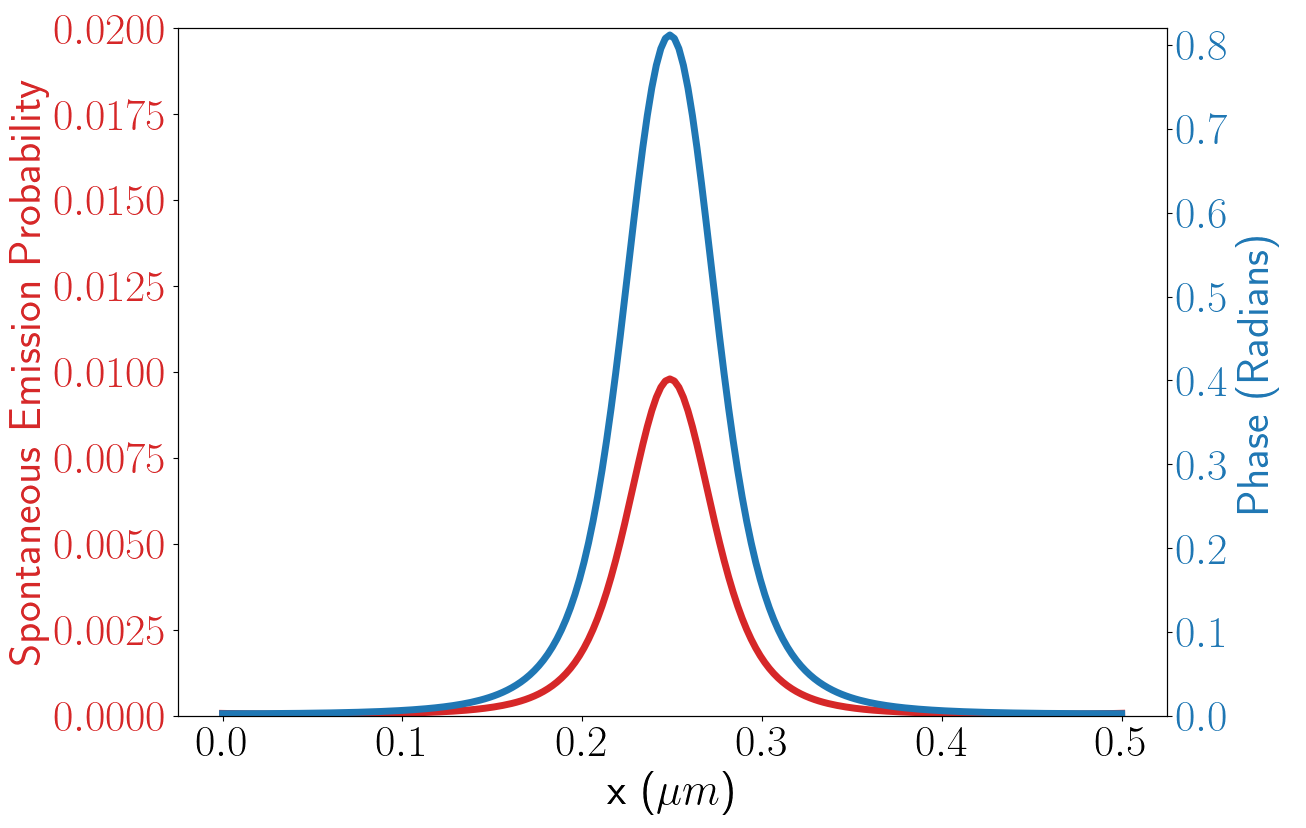}
	\caption{\label{fig:SEProb} The total incoherent spontaneous emission probability during the duration of the applied phase gate (solid red line) as well as the accumulated phase in radians (solid blue line) with respect to position. At the position of the target atom there is 0.82 radians phase accumulation while the spontaneous emission probability is kept $\approx0.01$. Lower spontaneous emission rates can be achieved by increasing the detuning, $\Delta$, of the Stark-shift laser from the excited level (at the expense of an increase in its Rabi frequency so that the applied phase $\Omega_{stark}^2/(2 \Delta)  T $ remains unchanged)}
\end{figure}

For the numerical simulations of Fig.~\ref{fig:PhaseCrossSection}, the spontaneous emission rate due to the non-adiabatic corrections to the dark state is negligible (at the level of 0.1\%). This is because of the high values for the probe and coupling laser Rabi frequencies. The spontaneous emission rate is instead limited by the excitation to the radiating level (followed by incoherent spontaneous emission) due to the Stark shift laser. In Fig.~\ref{fig:SEProb}, we plot the total spontaneous emission probability during the whole time duration of the applied phase gate, as a function of position for the conditions of the third inset of Fig.~\ref{fig:PhaseCrossSection} (i.e., for a coupling laser Rabi frequency of $\Omega_{C,max}=208 \Gamma_a$ at the intensity maximum). Here, we calculate the total spontaneous emission probability is calculated for each spatial point as $\int_0^T \rho_{44} \Gamma_a dt$ (the integration is over the whole time duration of the simulation). For completeness, we also plot the applied phase (in linear scale) as a function of position. At the intensity minimum of the coupling laser (i.e., at the position of the addressed qubit), the applied phase is about $\pi/4$, with a spontaneous emission probability of $approx0.01$. Lower spontaneous emission rates can be achieved by increasing the detuning, $\Delta$, of the Stark-shift laser from the excited level (at the expense of an increase in its Rabi frequency so that the applied phase $\Omega_{stark}^2/(2 \Delta)  T $ remains unchanged).

\section{Conclusions}

Neutral atoms have made great strides over the last decade towards a scalable quantum computing architecture.  Despite this great progress, there are still outstanding challenges that need to be overcome in neutral atom quantum computing. Even with high fidelity gates,  implementation of cross-talk free qubit measurements, which are required for error correction, is an outstanding challenge for neutral atom qubits. 
Another challenge in neutral atom arrays is the required high optical power of the trapping light.  To address these challenges, we have proposed to integrate dark-state based localization techniques into a neutral atom quantum computing architecture and suggested two schemes, one for state-projective measurement, and other for single-qubit phase gates. Density-matrix numerical simulations in $^{87}$Rb atoms show that both approaches can achieve a spatial resolution well into the nanoscale regime.  

\section{Acknowledgements}

We thank Diptaranjan Das for many helpful discussions. This work was was supported by National Science Foundation (NSF) Grant No. 2016136 for the QLCI center Hybrid Quantum Architectures and Networks and also by the Vilas Associates Award of the University of Wisconsin-Madison.  The simulations that are reported in this work were performed using the Computer Cluster at UW-Madison's Center for High Throughput Computing (CHTC). 

\clearpage

\begin{appendices}

\section{Density Matrix Equations
}\label{sec:densityMatrix}

For the four level EIT schemes that were discussed above, we numerically simulate the systems by calculating the time evolution of the $4\cross4$ density matrix $\rho$.  The Von Neumann equation for the density matrix $\rho$, without any relaxation, is~\cite{proposal_63_scully},

\begin{equation}
\label{eq:VonNeumann}
    \dot{\rho}=-\frac{i}{\hbar}\left[\mathcal{H},\rho \right].
\end{equation}

In order to include decay and dephasing processes, we add the relaxation matrix $\Gamma$ with the following matrix elements into the Von Neumann equation:
\hfill 
\begin{equation}
    \bra{n}\Gamma \ket{m}=\gamma_n \delta_{mn}.
\end{equation}
\hfill 
The total equation of motion with the relaxation matrix of above is:
\hfill 
\begin{equation}
    \dot{\rho}=-\frac{i}{\hbar}\left[\mathcal{H},\rho \right]-\frac{1}{2} \{ \Gamma,\rho\}.
\end{equation}
\hfill 

\noindent In order to simplify the notation that will follow, for the four level scheme of Fig.~\ref{fig:fourLevel}, we define the states $\ket{F=1, m_F=-1}\rightarrow \ket{a}$, $\ket{F=1, m_F=0}\rightarrow \ket{b}$, $\ket{F=1, m_F=-1}\rightarrow \ket{c}$ and $\ket{F'=0, m_F'=0}\rightarrow \ket{e}$. We define the relevant energies for these states to be $\hbar \omega_a$, $\hbar \omega_b$, $\hbar \omega_c$, and $\hbar \omega_e$, respectively. The states $\ket{a}$ and $\ket{c}$ are coupled to the excited state $\ket{e}$ with a coupling laser with a Rabi frequency of $\Omega_C$ and $\ket{b}$ state is coupled to the excited state $\ket{e}$ with a probe laser with the Rabi frequency $\Omega_P$. The unperturbed (zero field) Hamiltonian for the four atomic levels is:

\begin{equation}
H_0=
    \begin{bmatrix}
    \hbar \omega_a&0&0&0\\
    0&\hbar \omega_b&0&0\\
    0&0&\hbar \omega_c&0\\
    0&0&0&\hbar \omega_e\\
    \end{bmatrix}.
\end{equation}

 \noindent The dipole interaction Hamiltonian, that describes the interaction of the four levels  with the probe and coupling lasers ( frequencies $\Omega_C$ and $\Omega_P$) is:

\begin{equation}
\Vec{\mu} \cdot \Vec{E}=-\frac{1}{2}
    \begin{bmatrix}
    0&0&0&\Omega_C e^{-i \omega_C t}\\
    0&0&0&\Omega_P e^{-i \omega_P t}\\
    0&0&0&\Omega_C e^{-i \omega_C t}\\
    \Omega_C e^{i \omega_C t}&\Omega_P e^{i \omega_P t}&\Omega_C e^{i \omega_C t}&0\\
    \end{bmatrix}
\end{equation}

 \noindent Note that $\omega_P$ and $\omega_C$ are the phase differences on the rabi lasers. Whereas, $\omega_a$, $\omega_b$, $\omega_c$ and $\omega_e$ are the transition frequencies. When one combines the dipole interaction Hamiltonian and the unperturbed (zero-field) Hamiltonian, one gets the total Hamiltonian:

\begin{equation}
H=\frac{1}{2}
    \begin{bmatrix}
    \hbar \omega_a&0&0&-\Omega_C e^{-i \omega_C t}\\
    0&\hbar \omega_b&0&-\Omega_P e^{-i \omega_P t}\\
    0&0&\hbar \omega_c&-\Omega_C e^{-i \omega_C t}\\
    -\Omega_C e^{i \omega_C t}&-\Omega_P e^{i \omega_P t}&-\Omega_C e^{i \omega_C t}&\hbar \omega_e\\
    \end{bmatrix}.
\end{equation}

   \noindent We next use the Hamiltonian of Eq.~(12) in the Von Neumann equation of Eq.~(9) and write the differential equations for the elements of the density matrix $\rho(t)$. Below, the subscripts of $\rho$ are consistent the state labeling that was described in the previous paragraph. For example, the diagonal density matrix element for the state $\ket{a}$ is $\rho_{aa}$. The complete set of equations for the density matrix elements are:

\begin{equation*}
\begin{split}
        \dot{\rho}_{aa}&=\frac{i\Omega_C}{2}e^{i \omega_C t} \rho_{ae}^*-\frac{i\Omega_C}{2}e^{-i \omega_C t} \rho_{ae}-\Gamma_1 \rho_{aa}\\
        &+\frac{\Gamma_e \rho_{ee}}{3}\\
        \dot{\rho}_{ab}&=\frac{i\Omega_C}{2}e^{i \omega_C t} \rho_{be}^*-\frac{i\Omega_P}{2}e^{-i \omega_P t} \rho_{ae}\\
        &-\frac{\rho_{ab}}{2}(\Gamma_1+\Gamma_2)+i\rho_{ab}(\omega_b-\omega_a)\\
        \dot{\rho}_{ac}&=\frac{i\Omega_C}{2}e^{i \omega_C t} \rho_{ce}^*-\frac{i\Omega_C}{2}e^{-i \omega_C t} \rho_{ae}\\
        &-\frac{\rho_{ac}}{2}(\Gamma_a+\Gamma_{c})+i\rho_{ac}(\omega_{c}-\omega_a)\\
        \dot{\rho}_{ae}&=\frac{i\Omega_C}{2}e^{i \omega_C t} \rho_{ee}^*
        -\frac{i\Omega_C}{2}e^{-i \omega_C t} \rho_{aa}-\frac{i\Omega_P}{2}e^{-i \omega_P t}\rho_{ab}\\
        &-\frac{i\Omega_C}{2}e^{-i \omega_C t}\rho_{ac}-\frac{\rho_{ae}}{2}(\Gamma_a+\Gamma_e)\\
        &+i\rho_{ae}(\omega_e-\omega_a)\\
        \dot{\rho}_{bb}&=\frac{i\Omega_P}{2}e^{i \omega_P t} \rho_{be}^*-\frac{i\Omega_P}{2}e^{-i \omega_P t} \rho_{be}-\Gamma_2 \rho_{bb}\\
        &+\frac{\Gamma_e \rho_{ee}}{3}\\
    \end{split}
\end{equation*}

\begin{equation}
\begin{split}      
        \dot{\rho}_{bc}&=\frac{i\Omega_P}{2}e^{i \omega_P t} \rho_{ce}^*-\frac{i\Omega_C}{2}e^{-i \omega_C t} \rho_{be}\\
        &-\frac{\rho_{bc}}{2}(\Gamma_b+\Gamma_{c})+i\rho_{bc}(\omega_{c}-\omega_b)\\
        \dot{\rho}_{be}&=\frac{i\Omega_P}{2}e^{i \omega_P t} \rho_{ee}^*-\frac{i\Omega_C}{2}e^{-i \omega_C t} \rho_{ab}^*\\
        &-\frac{i\Omega_P}{2}e^{i \omega_P t} \rho_{bb} -\frac{i\Omega_C}{2}e^{i \omega_C t} \rho_{bc}-\frac{\rho_{be}}{2}(\Gamma_b+\Gamma_e)\\
        &+i\rho_{be}(\omega_e-\omega_b)\\
        \dot{\rho}_{cc}&=\frac{i\Omega_C}{2}e^{i \omega_C t} \rho_{ce}^*-\frac{i\Omega_C}{2}e^{-i \omega_C t} \rho_{ce}-\Gamma_{c} \rho_{cc}\\
        &+\frac{\Gamma_e \rho_{ee}}{3}\\
        \dot{\rho}_{ce}&=\frac{i\Omega_C}{2}e^{i \omega_C t} \rho_{ee}-\frac{i\Omega_C}{2}e^{-i \omega_C t} \rho_{ac}^*
        -\frac{i\Omega_P}{2}e^{i \omega_P t} \rho_{bc}^*\\
        &-\frac{i\Omega_C}{2}e^{i \omega_C t} \rho_{cc}-\frac{\rho_{ae}}{2}(\Gamma_a+\Gamma_e)\\
        &+i\rho_{ae}(\omega_e-\omega_a)\\
        \dot{\rho}_{ee}&=\frac{i\Omega_C}{2}e^{-i \omega_C t} \rho_{ae}+\frac{i\Omega_P}{2}e^{-i \omega_P t} \rho_{be}\\
        &+\frac{i\Omega_C}{2}e^{-i\Omega_C t} \rho_{ce}\\
        &-\frac{i\Omega_C}{2}\rho_{ae}^*-\frac{i\Omega_P}{2}\rho_{be}^*-\frac{i\Omega_C}{2}e^{i \omega_C t} \rho_{ce}^*-\Gamma_4\rho_{ee}\\
    \end{split}
\end{equation}

Where the terms $\rho_{ii}$s define the relevant density matrix element of $\rho$. When one transforms to the rotating frame, the terms transform to

\begin{equation}
\begin{split}
  \Tilde{\rho}_{ae}&=\rho_{ae}e^{-i\omega_C t}\\
  \Tilde{\rho}_{be}&=\rho_{be}e^{-i\omega_P t}\\
  \Tilde{\rho}_{ce}&=\rho_{ce}e^{i\omega_C t}\\
  \Tilde{\rho}_{ab}&=\rho_{ab}e^{-i(\omega_C-\omega_P )t}\\
  \Tilde{\rho}_{bc}&=\rho_{bc}e^{i(\omega_C-\omega_P )t}.\\
\end{split}
\end{equation}

Thus, the differential equations  for the terms of the density matrix can be updated as,

\begin{equation*}
\begin{split}
    \Dot{\Tilde{\rho}}_{aa}&=\frac{i\Omega_C}{2}\Tilde{\rho}_{ae}^*-\frac{i\Omega_C}{2}\Tilde{\rho}_{ae}-\Gamma_1\Tilde{\rho}_{aa}+\frac{\Gamma_e \Tilde{\rho}_{ee}}{3}\\
    \Dot{\Tilde{\rho}}_{ab}&=\frac{i\Omega_C}{2}\Tilde{\rho}_{be}^*-\frac{i\Omega_P}{2}\Tilde{\rho}_{ae}-\frac{\Tilde{\rho}_{ab}}{2}(\Gamma_a+\Gamma_{c})\\
    &+i\Tilde{\rho}_{ab}(\omega_b-\omega_a-\omega_C+\omega_P)\\
    \Dot{\Tilde{\rho}}_{ac}&=\frac{i\Omega_C}{2}\Tilde{\rho}_{ce}^*-\frac{i\Omega_C}{2}\Tilde{\rho}_{ab}-\frac{\Tilde{\rho}_{ac}}{2}(\Gamma_a+\Gamma_{c})\\
    &+i\Tilde{\rho}_{ac}(\omega_{c}-\omega_a)\\
    \Dot{\Tilde{\rho}}_{ae}&=\frac{i\Omega_C}{2}\Tilde{\rho}_{ee}-\frac{i\Omega_C}{2}\Tilde{\rho}_{aa}-\frac{i\Omega_P}{2}\Tilde{\rho}_{ab}-\frac{i\Omega_C}{2}\Tilde{ac}\\
    &-\frac{\Tilde{\rho}_{ae}}{2}(\Gamma_a+\Gamma_e)+i\Tilde{\rho}_{ae}(\omega_e-\omega_a-\omega_b)\\
    \Dot{\Tilde{\rho}}_{bb}&=\frac{i\Omega_P}{2}\Tilde{\rho}_{be}^*-\frac{i\Omega_P}{2}\Tilde{\rho}_{be}-\Gamma_2\Tilde{\rho}_{bb}\\
    &+\frac{\Gamma_4\Tilde{\rho}_{ee}}{2}(\Gamma_a+\Gamma_e)\\
\end{split}
\end{equation*}
\begin{equation}
\begin{split}   
     \Dot{\Tilde{\rho}}_{bc}&=\frac{i\Omega_P}{2}\Tilde{\rho}_{ce}^*-\frac{i\Omega_C}{2}\Tilde{\rho}_{ce}-\frac{\Tilde{\rho}_{bc}}{2}(\Gamma_b+\Gamma_{c})\\
     &+i\Tilde{\rho}_{bc}(\omega_{c}-\omega_b-\omega_P+\omega_C)\\
     \Dot{\Tilde{\rho}}_{be}&=\frac{i\Omega_P}{2}\Tilde{\rho}_{ee}-\frac{i\Omega_C}{2}\Tilde{\rho}_{ab}^*-\frac{i\Omega_P}{2}\Tilde{\rho}_{bb}-\frac{i\Omega_C}{2}\tilde{\rho}_{bc}\\
     &-\frac{\Tilde{\rho}_{be}}{2}(\Gamma_b+\Gamma_e)+i\Tilde{\rho}_{be}(\omega_e-\omega_b-\omega_P)\\
     \Dot{\Tilde{\rho}}_{cc}&=\frac{i\Omega_C}{2}\Tilde{\rho}_{ce}^*-\frac{i\Omega_C}{2}\Tilde{\rho}_{ce}-\Gamma_{3c}\Tilde{\rho}_{cc}+\frac{\Gamma_e\Tilde{\rho}_{ee}}{3}\\
     \Dot{\Tilde{\rho}}_{ce}&=\frac{i\Omega_C}{2}\Tilde{\rho}_{ee}-\frac{i\Omega_C}{2}\Tilde{\rho}_{ac}^*-\frac{i\Omega_P}{2}\Tilde{\rho}_{bc}^*-\frac{i\Omega_C}{2}\Tilde{\rho}_{cc}\\
     &-\frac{\Tilde{\rho}_{ce}}{2}(\Gamma_{c}+\Gamma_e)+i\Tilde{\rho}_{ce}(\omega_e-\omega_{c}-\omega_C)\\
     \dot{\Tilde{\rho}}_{ee}&=\frac{i\Omega_C}{2}\Tilde{\rho}_{ae}+\frac{i\Omega_P}{2}\Tilde{\rho}_{be}+\frac{i\Omega_C}{2}\Tilde{\rho}_{ce}\\
     &-\frac{i\Omega_C}{2}\Tilde{\rho}_{ae}^*-\frac{i\Omega_P}{2}\Tilde{\rho}_{be}^*-\frac{i\Omega_C}{2}\Tilde{\rho}_{ce}^*-\Gamma_4\Tilde{\rho}_{ee}.\\
    \end{split}
\end{equation}

Since the energies of $\ket{a}$ and $\ket{c}$ are the same,  $\omega_a=\omega_{c}$ holds true. Thus, we define the decouplings,

\begin{equation}
    \begin{split}
    \Delta_{1}&=\omega_e-\omega_a-\omega_C\\
    \Delta_2&=\omega_e-\omega_b-\omega_P\\
    \end{split}
\end{equation}

Considering these decouplings, our differential equations become,

\begin{equation*}
\begin{split}
    \Dot{\Tilde{\rho}}_{aa}&=\frac{i\Omega_C}{2}\Tilde{\rho}_{ae}^*-\frac{i\Omega_C}{2}\Tilde{\rho}_{ae}-\Gamma_1\Tilde{\rho}_{aa}+\frac{\Gamma_e \Tilde{\rho}_{ee}}{3}\\
    \Dot{\Tilde{\rho}}_{ab}&=\frac{i\Omega_C}{2}\Tilde{\rho}_{be}^*-\frac{i\Omega_P}{2}\Tilde{\rho}_{ae}-\frac{\Tilde{\rho}_{ab}}{2}(\Gamma_a+\Gamma_{c})\\
    &+i\Tilde{\rho}_{ab}(\Delta_1-\Delta_2)\\
    \dot{\Tilde{\rho}}_{ac}&=\frac{i\Omega_C}{2}\Tilde{\rho}_{ce}^*-\frac{i\Omega_C}{2}\Tilde{\rho}_{ab}-\frac{\Tilde{\rho}_{ac}}{2}(\Gamma_a+\Gamma_{c})\\
    \Dot{\Tilde{\rho}}_{ae}&=\frac{i\Omega_C}{2}\Tilde{\rho}_{ee}-\frac{i\Omega_C}{2}\Tilde{\rho}_{aa}-\frac{i\Omega_P}{2}\Tilde{\rho}_{ab}-\frac{i\Omega_C}{2}\Tilde{ac}\\
    &-\frac{\Tilde{\rho}_{ae}}{2}(\Gamma_a+\Gamma_e)+i\Tilde{\rho}_{ae}(\omega_e-\omega_a-\omega_b)\\
    \dot{\Tilde{\rho}}_{bb}&=\frac{i\Omega_P}{2}\Tilde{\rho}_{be}^*-\frac{i\Omega_P}{2}\Tilde{\rho}_{be}-\Gamma_2\Tilde{\rho}_{bb}+\frac{\Gamma_4\Tilde{\rho}_{ee}}{2}(\Gamma_a+\Gamma_e)\\
    \end{split}
\end{equation*}
\begin{equation}
\begin{split}
     \dot{\Tilde{\rho}}_{bc}&=\frac{i\Omega_P}{2}\Tilde{\rho}_{ce}^*-\frac{i\Omega_C}{2}\Tilde{\rho}_{ce}-\frac{\Tilde{\rho}_{bc}}{2}(\Gamma_b+\Gamma_{c})\\
     &+i\Tilde{\rho}_{bc}(\Delta_2-\Delta_1)\\
     \dot{\Tilde{\rho}}_{be}&=\frac{i\Omega_P}{2}\Tilde{\rho}_{ee}-\frac{i\Omega_C}{2}\Tilde{\rho}_{ab}^*-\frac{i\Omega_P}{2}\Tilde{\rho}_{bb}-\frac{i\Omega_C}{2}\tilde{\rho}_{bc}\\
     &-\frac{\Tilde{\rho}_{be}}{2}(\Gamma_b+\Gamma_e)+i\Tilde{\rho}_{be}(\omega_e-\omega_b-\omega_P)\\
     \dot{\Tilde{\rho}}_{cc}&=\frac{i\Omega_C}{2}\Tilde{\rho}_{ce}^*-\frac{i\Omega_C}{2}\Tilde{\rho}_{ce}-\Gamma_{3c}\Tilde{\rho}_{cc}+\frac{\Gamma_e\Tilde{\rho}_{ee}}{3}\\
     \dot{\Tilde{\rho}}_{ce}&=\frac{i\Omega_C}{2}\Tilde{\rho}_{ee}-\frac{i\Omega_C}{2}\Tilde{\rho}_{ac}^*-\frac{i\Omega_P}{2}\Tilde{\rho}_{bc}^*-\frac{i\Omega_C}{2}\Tilde{\rho}_{cc}\\
     &-\frac{\Tilde{\rho}_{ce}}{2}(\Gamma_{c}+\Gamma_e)+i\Tilde{\rho}_{ce}(\omega_e-\omega_{c}-\omega_C)\\
     \dot{\Tilde{\rho}}_{ee}&=\frac{i\Omega_C}{2}\Tilde{\rho}_{ae}+\frac{i\Omega_P}{2}\Tilde{\rho}_{be}+\frac{i\Omega_C}{2}\Tilde{\rho}_{ce}-\frac{i\Omega_C}{2}\Tilde{\rho}_{ae}^*\\
     &-\frac{i\Omega_P}{2}\Tilde{\rho}_{be}^*-\frac{i\Omega_C}{2}\Tilde{\rho}_{ce}^*-\Gamma_4\Tilde{\rho}_{ee}.\\
    \end{split}
\end{equation}

\hfill

After getting the rotating-frame coupled differential equations, a fourth order Runge-Kutta Algorithm~\cite{Runge,Kutta} is used to numerically solve these equations. The solutions to these equations are used in Section~\ref{section:results}.

\end{appendices}
\hfill \newpage
\bibliographystyle{apsrev4-2}
\typeout{}
\bibliography{ref}

\end{document}